\documentclass{ifacconf}

\usepackage{natbib}        
\usepackage{amsmath, amssymb,amsfonts, lipsum, mathtools, cuted, color,thmtools}
\usepackage{graphicx}

\allowdisplaybreaks
\begin{document}
	\begin{frontmatter}
		
		\title{Linear Fault Estimators for Nonlinear Systems: An Ultra-Local Model Design\thanksref{footnoteinfo}} 
		
		\thanks[footnoteinfo]{This publication is part of the project Digital Twin project
			4.3 with project number P18-03 of the research programme
			Perspectief which is (mainly) financed by the Dutch Research
			Council (NWO).}
		
		\author[tue]{Farhad Ghanipoor}
		\author[tue]{Carlos Murguia}
		\author[delft]{Peyman Mohajerin Esfahani}
		\author[tue]{Nathan van de Wouw}
		
		\address[tue]{Mechanical Engineering Department, Eindhoven University of Technology, Eindhoven, The Netherlands (e-mails: f.ghanipoor@tue.nl, C.G.Murguia@tue.nl, and N.v.d.Wouw@tue.nl).}  
		\address[delft]{Delft Center for Systems and Control, Delft University of Technology, Delft, The Netherlands (e-mail: P.MohajerinEsfahani@tudelft.nl).}             

		\begin{abstract}
			This paper addresses the problem of robust process and sensor fault reconstruction for nonlinear systems. The proposed method augments the system dynamics with an approximated internal linear model of the combined contribution of \emph{known} nonlinearities and \emph{unknown} faults -- leading to an approximated linear model in the augmented state. We exploit the broad modeling power of ultra-local models to characterize this internal dynamics. We use a linear filter to reconstruct the augmented state (simultaneously estimating the state of the original system and the sum of nonlinearities and faults). Having this combined estimate, we can simply subtract the analytic expression of nonlinearities from that of the corresponding estimate to reconstruct the fault vector. Because the nonlinearity does not play a role in the filter dynamics (it is only used as a static nonlinear output to estimate the fault), we can avoid standard restrictive assumptions like globally (one-sided) Lipschitz nonlinearities and/or the need for Lipschitz constants to carry out the filter design. The filter synthesis is posed as a mixed $H_2/H_{\infty}$ optimization problem where the effect of disturbances and model mismatches is minimized in the $H_{\infty}$ sense, for an acceptable $H_2$ performance with respect to measurement noise.
		\end{abstract}
		
	\end{frontmatter}
	
	\section{Introduction} \label{sec:introduction}
	Predictive maintenance is a vital technology that guarantees the reliable operation of high-tech industrial systems. A fundamental element for predictive maintenance is fault estimation (for severity assessment) \cite{ghanipoor2022ultra}. If we have an accurate estimate of the fault effect in the system, we can predict the magnitude of the slowly increasing fault signal and schedule predictive maintenance actions accordingly. Many fault estimation methods have been developed for linear dynamical systems (see, e.g., \cite{liu2013sensor,liu2012fuzzy} for results on linear stochastic and switching systems). However, most engineered systems (e.g., robotics, power/water networks, transportation, and manufacturing) are highly nonlinear in nature. Methods for nonlinear systems are still under development, see, e.g.,  \cite{esfahani2015tractable}, \cite{de2001geometric}, \cite{pan2021dynamic}, \cite{lan2020robust}, and \cite{Adaptive}. In \cite{Adaptive}, the authors address the problem for nonlinear systems with uniformly Lipschitz nonlinearities, process faults only (i.e., no sensor faults), and assume the so-called matching condition (the rank of the fault distribution matrix is invariant under left multiplication by the output matrix) is satisfied. An adaptive filter is introduced that approximately reconstructs the actuator fault vector in this configuration. Although the matching condition makes the problem tractable, it significantly reduces the class of systems that can benefit from the results. In \cite{Phan2021}, using Nonlinear Unknown Input Observers (NUIO), adaptive Radial Basis Function Neural Networks (RBFNN), and assuming the matching condition, a fault reconstruction scheme is provided for both sensor and process faults. The authors prove the bounded fault estimation errors of the provided scheme.
	
	In \cite{zhu2015fault}, the requirement for matching condition is relaxed. The authors consider simultaneous sensor and process faults, Lipschitz nonlinearities, and use a standard fault observability condition for the linear part of the dynamics, \cite{Patton_Input_Observability}. The problem is addressed through the concept of intermediate observers, which involve two specialized observers: one that estimates the fault and the other the state. Their scheme guarantees bounded fault estimation errors. In \cite{van2022multiple}, simultaneous additive and multiplicative process faults are considered in the scope of discrete-time system models. Herein, the fault estimation problem is addressed by decoupling process nonlinearities and disturbances from the estimation filter dynamics and using regression techniques to approximately estimate fault signals. Linear filters can be obtained by this decoupling nonlinearities, allowing linear techniques to reconstruct fault signals. Nonetheless, the system dynamics are subject to stringent conditions due to decoupling conditions, resulting in limited practicality of these findings. 
	
	It should be noted that the results discussed above regarding nonlinear systems provide only an \textit{approximate} reconstruction of fault vectors. This means that they guarantee bounded estimation errors, which, if within a certain range, can still provide a reliable estimate of the actual fault. The absence of internal models for fault signals creates a difficulty in achieving perfect fault estimation with zero errors. In \cite{ghanipoor2022ultra}, an internal model for the fault vector is used, and this allows to guarantee zero error in the absence of external disturbances and noise for some classes of faults. However, this solution only applies to systems with globally Lipschitz nonlinearities and the effect of measurement noise and exogenous disturbances is not addressed. 
	We know that some of the industrial systems does not satisfy globally Lipschitz condition such as robotic manipulators with Euler-Lagrange equations. 
	
	Considering the discussed literature, the main contribution of this manuscript is a fault estimation scheme for process and sensor faults that \emph{does not require} globally Lipschitz nonlinearities, known Lipschitz constants, and the matching condition. This scheme augments the system dynamics with an approximated internal linear model of the combined contribution of \emph{known} nonlinearities and \emph{unknown} faults -- leading to an approximated linear model in the augmented state. We exploit the broad modeling power of \emph{ultra-local models}, see \cite{Flies_Ultra_Local} and \cite{sira2018active} -- which refers to a class of phenomenological models that are only valid for very short time intervals -- to characterize this internal dynamics. We use a linear filter to reconstruct the augmented state (simultaneously estimating the state of the original system and the sum of nonlinearities and faults). If this estimate is sufficiently accurate, we can simply subtract the analytic expression of nonlinearities from that of the corresponding estimate to reconstruct the fault vector. Because the nonlinearity does not play a role in the filter dynamics (it is only used as a static nonlinear output to estimate the fault), we can avoid standard restrictive assumptions like globally (one-sided) Lipschitz nonlinearities and/or the need of Lipschitz constants to carry out the filter design. The filter synthesis is posed as a mixed $H_2/H_{\infty}$ semi-definite program where the effect of exogenous disturbances and the mismatch between ultra-local models and true internal dynamics is minimized in the $H_{\infty}$ sense, for an acceptable $H_2$ performance with respect to measurement noise. 
	
	
	\textbf{Notation:}
	The symbol $\mathbb{R}^{+}$ denotes the set of nonnegative real numbers. The $n \times n$ identity matrix is denoted by $I_n$	or simply $I$ if $n$ is clear
	from the context. Similarly, $n \times m$ matrices composed of
	only zeros are denoted by $\boldsymbol{0}_{n \times m}$ or simply $\boldsymbol{0}$ when their dimensions are clear. For positive definite (semi-definite) matrices, we use the notation $P \succ 0$ $(P \succeq 0)$. For negative definite (semi-definite) matrices, we use the notation $P \prec 0$ $(P \preceq 0)$. The notation $\text{col}[x_1, \ldots , x_n]$ stands for the column vector composed of the elements $x_1,\ldots,x_n$. This notation is also used when the components $x_i$ are vectors. The $\ell_2$ vector norm (Euclidean norm) and the matrix norm induced by the $\ell_2$ vector norm are both denoted as $||\cdot||$. For a vector-valued signal $f(t)$ defined for all $t \geq 0$, $||f(t)||_\infty := \sup_{t \geq 0} ||f(t)||$. For a differentiable function $W: \mathbb{R}^{n}  \to \mathbb{R}$ we denote by $\frac{\partial W}{\partial e}$ the row-vector of partial derivatives and by $\dot{W}(e)$ the total derivative of $W(e)$ with respect to time. For the transfer function $T(s)$, $\sigma_{\max }(T (s))$ shows the maximum singular value, and $T^H(s)$ the Hermitian transpose. We often omit time dependencies for notation simplicity.

	%


	
	\section{Problem Formulation} \label{sec:problem formulation}
	Consider the nonlinear system
	\begin{equation} \label{eq:sys}
		\left\{\begin{aligned}
			\dot{x} =& A x+ B u +  S g(V x, u, t) + D \omega + F_{x} f_x(x, u, t) ,\\
			y=& C x + F_y f_y(x, t) + \nu,
		\end{aligned}\right.
	\end{equation}
	where $t \in \mathbb{R}^{+}$, $x \in {\mathbb{R}^{n}}$, $y \in {\mathbb{R}^{{m}}}$, and $u \in {\mathbb{R}^{{l}}}$ are time, state, measured output, and known input vectors, respectively. Function $g: \mathbb{R}^{n_v} \times \mathbb{R}^{l} \times \mathbb{R}^{+} \to \mathbb{R}^{n_{g}}$ is a nonlinear \emph{known} vector field. Vectors $\omega: \mathbb{R}^{+}  \to \mathbb{R}^{n_d}$ and $\nu: \mathbb{R}^{+}  \to \mathbb{R}^{m}$ are unknown bounded external disturbances and measurement noise, respectively. Functions $f_x: \mathbb{R}^{n} \times \mathbb{R}^{l} \times \mathbb{R}^{+} \to \mathbb{R}^{n_{f_x}}$ and $f_y: \mathbb{R}^{n} \times \mathbb{R}^{+} \to \mathbb{R}^{n_{f_y}}$ denote \emph{unknown} process and sensor fault vectors, respectively. Matrices $(A, B, S, V, C, F_x, F_y)$ are of appropriate dimensions and ${n}, {m}, l, n_g, n_{v}, n_{d}, n_{f_x}, n_{f_y} \in \mathbb{N}$. We use matrix $S$ to indicate in what equations the nonlinearity $g(Vx,u,t)$ appears explicitly, and $V$ is used to specify which states play a role in the known nonlinearity. Matrices $F_x$ and $F_y$ denote process and sensor fault distribution matrices, respectively. 
	
	Let $S^{\dagger}$ denote the Moore-Penrose pseudo-inverse of matrix $S$, and define  $Q_1 R_1 := S^{\dagger} F_x$ and $Q_2 R_2 := (I-SS^{\dagger}) F_x$, where $Q_i$ ($R_i$) denotes a full column (row) rank matrix of appropriate dimensions, $i=1,2$. Note that this factorization always exists (rank decomposition). Then, we can rewrite the nonlinear system in \eqref{eq:sys} as follows:
	\begin{subequations} \label{eq:sys_transformed}
		\begin{equation} \label{eq:sys_transformed_ss}
			\left\{\begin{aligned}
				\dot{x} =& A x+ B u +  S \big( g(V x, u, t) + Q_1 f_n(x, u, t)\big)  \\[1mm]
				& + Q_2 f_l(x, u, t) + D \omega,\\[1.5mm]
				y=& C x + F_y f_y(x, t) + \nu,
			\end{aligned}\right. 			\\[1mm]
		\end{equation}\\[-6mm]
		\begin{equation} \label{eq:fault_change}
			\Bigg\{
			\begin{array}{r@{}l}
				f_n &{}:= R_1 f_x,\\[1mm]
				f_l &{}:= R_2 f_x.
			\end{array}
			\hspace{49.5mm}
		\end{equation}
	\end{subequations}
	This transformation does not lead to loss of generality and will play an instrumental role to the fault estimation algorithm that we propose. 
	
	\begin{assum}\emph{\textbf{(State and Input Boundedness)}  The state variable $x$ and the input $u$ remain bounded in some compact region of interest $\mathcal{D}=\mathcal{D}_{x} \times \mathcal{D}_{u} \subset \mathbb{R}^{n} \times \mathbb{R}^{l}$, before and after the occurrence of a fault.}
		\label{assum:state_boundedness}
	\end{assum}

	
	\begin{assum}\emph{\textbf{($\mathcal{C}^r$ Fault and Nonlinearity)} Vectors $g(V x, u, t) + Q_1 f_n(x, u,t)$, $f_l(x, u, t)$, and $f_y(x, t)$ are $r$ times differentiable with respect to $t$, and their $r$-th derivative is uniformly bounded.}
		\label{assum:cr_assumption}
	\end{assum}
	
	\begin{assum}\emph{\textbf{($\mathcal{C}^1$ Measurement Noise)}} The measurement noise vector $\nu(t)$ in \eqref{eq:sys} is bounded uniformly in $t$ and differentiable, i.e., the total derivative with respect to time $\dot{\nu}(t)$ exists, is continuous, and bounded uniformly in $t$.
		\label{assum:noise_diff}
	\end{assum}

	\subsection{Ultra Local Model}
	
	Define the functions $\beta_1(x,u,t) := g(V x,u,t) + Q_1 f_n(x,u,t)$, $\beta_2(x,u,t) := f_l(x,u,t)$, and $\beta_3(x,t) := f_y(x,t)$. We refer to these $\beta_k(t)$ as the (fault-induced) lumped disturbances. Under Assumption \ref{assum:cr_assumption}, using the notion of ultra-local models, \cite{Flies_Ultra_Local}, we can write linear approximated internal models of $\beta_k(t)$. Consider the approximated model:
	\begin{equation}\label{eq:ultra_local_internal_model}
		\left\{\begin{aligned}
			\dot{\zeta}_{j_k}(t) &= \zeta_{{(j+1)}_k}(t), \qquad &&j_k \in \{1, \dots, r_k-1\},\\[.25mm]
			\dot{\zeta}_{r_k}(t) &= \mathbf{0}, \qquad &&k \in \{1, 2, 3\},\\[.25mm]
			\bar\beta_k(t) &= \zeta_{1_k}(t),
		\end{aligned}\right.
	\end{equation}
	where $\bar\beta_k(t), k \in \{1, 2, 3\}$, is an approximation of the actual $\beta_k(t)$ introduced above. Note that each of these models corresponds to an entry-wise $r_k$-th order Taylor polynomial-in-time approximation at time $t$ of $\beta_k(t)$. The accuracy of the approximated model increases as $d\beta_k(t)/dt^{r_k}$ goes to zero (entry-wise) and it is exact for $d\beta_k(t)/dt^{r_k} = \mathbf{0}$. To design such an observer, we extend the system state, $x(t)$, with the states of the internal system (not model) for each of the lumped disturbances where we consider $\dot{\zeta}_{r_k}(t) = d\beta_k(t)/dt^{r_k}$, $k \in \{1, 2, 3\}$ to consider actual system of the model provided for the lumped disturbances, augment it with the system dynamics in \eqref{eq:sys_transformed}. We then design a linear observer for the augmented linearized system to simultaneously estimate $x(t)$ and $\beta_k(\cdot)$. We remark that the number of the derivatives, $r_k$, for each of the above-mentioned models in \eqref{eq:ultra_local_internal_model} is problem-dependent, see \cite{ghanipoor2022ultra} for discussion on selection of  $r_k$. 
	
	\subsection{Augmented Dynamics}
	Define $\rho_k:=\text{col}[\dot{\beta_k},\ldots,\beta_k^{(r_k-1)}]$, $k \in \{1, 2, 3\}$ and the augmented state $x_a:=\text{col}[x,\beta_1, \rho_1,\beta_2, \rho_2,\beta_3, \rho_3]$, and write the augmented linear dynamics using \eqref{eq:sys_transformed}-\eqref{eq:ultra_local_internal_model} as
	\begin{subequations} 		\label{eq:augmented}
		\begin{equation}\label{eq:augmented_system}
			\begin{aligned}
				\left\{\begin{aligned}
					\dot{x}_{a} &=A_{a} x_{a}+B_{a} u +D_{a} \omega_{a},\\
					y &= C_{a} x_{a}+\nu,
				\end{aligned}\right.
			\end{aligned}
		\end{equation}
		\begin{equation} 		\label{eq:augmented_matrices}
			\begin{aligned}
				A_{a} &:=\left[\begin{array}{ccccccccccc}
					A & S & \boldsymbol{0}  & Q_2 & \boldsymbol{0} & \boldsymbol{0} & \boldsymbol{0} \\
					\boldsymbol{0} & \boldsymbol{0} & I_{d_1} & \boldsymbol{0}  & \boldsymbol{0} & \boldsymbol{0}  & \boldsymbol{0} \\
					\boldsymbol{0} & \boldsymbol{0} & \boldsymbol{0} & \boldsymbol{0} & \boldsymbol{0} & \boldsymbol{0} & \boldsymbol{0} \\
					\boldsymbol{0} & \boldsymbol{0} & \boldsymbol{0} & \boldsymbol{0}  & I_{d_2}  & \boldsymbol{0}  & \boldsymbol{0} \\
					\boldsymbol{0} & \boldsymbol{0} & \boldsymbol{0} & \boldsymbol{0} & \boldsymbol{0} & \boldsymbol{0} & \boldsymbol{0} \\
					\boldsymbol{0} & \boldsymbol{0} & \boldsymbol{0} & \boldsymbol{0}  & \boldsymbol{0} & \boldsymbol{0} & I_{d_3} \\
					\boldsymbol{0} & \boldsymbol{0} & \boldsymbol{0} & \boldsymbol{0} & \boldsymbol{0} & \boldsymbol{0} & \boldsymbol{0} \\
				\end{array}\right],  
							D_{a} := \left[\begin{array}{cccc}
									D & \boldsymbol{0} & \boldsymbol{0}  & \boldsymbol{0} \\
									\boldsymbol{0} & \boldsymbol{0} & \boldsymbol{0}  & \boldsymbol{0} \\
									\boldsymbol{0} & I_{d_1} & \boldsymbol{0}  & \boldsymbol{0} \\
									\boldsymbol{0} & \boldsymbol{0} & \boldsymbol{0}  & \boldsymbol{0} \\
									\boldsymbol{0} & \boldsymbol{0} & I_{d_2}  & \boldsymbol{0} \\
									\boldsymbol{0} & \boldsymbol{0} & \boldsymbol{0}  & \boldsymbol{0} \\
									\boldsymbol{0} & \boldsymbol{0} & \boldsymbol{0}  & I_{d_3} \\
								\end{array}\right] \\
				B_a &:=	\left[\begin{array}{ll}
					B^\top &
					\boldsymbol{0}
				\end{array}\right]^\top, \quad
				C_a := \left[\begin{array}{lllllll}	C & \boldsymbol{0} & \boldsymbol{0} & \boldsymbol{0} & \boldsymbol{0} & F_y & \boldsymbol{0} \end{array}\right] \\
				\omega_{a} &:=	\left[\begin{array}{llll}
					\omega(t)^\top &
					\beta_1^{{(r_1)}} (\cdot)^\top &
					\beta_2^{{(r_2)}} (\cdot)^\top &
					\beta_3^{{(r_3)}} (\cdot)^\top 
				\end{array}\right]^\top
			\end{aligned}
		\end{equation}
	\end{subequations}
	with $d_1 = (r_1-1)n_{g}, d_2= (r_2-1)n_{f_l}$, $n_{f_l}$ being the number of rows of $R_2$ in \eqref{eq:fault_change} and $d_3 = (r_3-1)n_{f_y}$.
	\subsection{Linear Joint State-Fault Estimator}
	We propose the following linear observer to estimate the fault vector:
	\begin{subequations} \label{eq:observer}
		\begin{equation}
			\begin{aligned} 		\label{eq:observer_dynamics}
				\dot{z} &=N z+G u+L y, \\
				\hat{x}_a &=z-E y,
			\end{aligned}
		\end{equation}
		with observer state $z \in {\mathbb{R}^{n_z}}$, $n_z = n+r_1 n_{g}+r_2 n_{f_l}+r_3 n_{f_y}$, estimate $\hat{x}_a$, and matrices $(N,G,L)$ defined as
		\begin{equation}		\label{eq:observer_matrices}
			\begin{aligned}
				N &:=M A_a-K C_a,
				&M&:=I+E C_a, \\[1mm]
				G&:=M B_a,
				&L&:=K(I+C_a E)-M A_a E.
			\end{aligned}
		\end{equation}
	\end{subequations}
	Matrices $E$ and $K$ are observer gains to be designed. Note that, given $\hat{x}_a$, fault estimates, $\hat{f}_x(\cdot)$ and $\hat{f}_y(\cdot)$, can be constructed as follows:
	\begin{equation} \label{eq:algebraic_fault_estimate}
		\begin{aligned}
			\hat{f}_{x}	&= \left[\begin{array}{c}
				R_1 \\
				R_2
			\end{array}\right]^L
			\left[\begin{array}{c}
				Q_1^L\big( \bar{C}_1 \hat{x}_{a}- g\left(V_a \hat{x}_{a}, u, t\right)\big) \\
				\bar{C}_2 \hat{x}_{a}
			\end{array}\right], \quad
			\hat{f}_{y}	=\bar{C}_3 \hat{x}_{a},
		\end{aligned}	
	\end{equation}
	where
	\begin{equation*}
		\begin{aligned}
			\bar{C}_1 &:= {\left[\begin{array}{ccccccc}
					\boldsymbol{0} & I_{n_{g}} & \boldsymbol{0} & \boldsymbol{0} & \boldsymbol{0} & \boldsymbol{0} & \boldsymbol{0}
				\end{array}\right]}, \quad
			\bar{C}_2 := {\left[\begin{array}{ccccccc}
					\boldsymbol{0} & \boldsymbol{0} & \boldsymbol{0} & I_{n_{f_l}} & \boldsymbol{0} & \boldsymbol{0} & \boldsymbol{0}
				\end{array}\right]}, \\
			\bar{C}_3 &:= {\left[\begin{array}{ccccccc}
					\boldsymbol{0} & \boldsymbol{0} & \boldsymbol{0} & \boldsymbol{0} & \boldsymbol{0} & I_{n_{f_y}} & \boldsymbol{0}
				\end{array}\right]}, \quad
			V_a := {\left[\begin{array}{cc}
					V & \boldsymbol{0}
				\end{array}\right]},
		\end{aligned}
	\end{equation*}
	and $[\cdot]^L$ denotes left inverse. Note that, by construction $[R_1^\top,R_2^\top]^\top$ and $Q_1^L$, are full column rank; hence, they are always left invertible. Note that we assume that the fault-free system dynamics is known. Hence, the nonlinear function $g(\cdot)$, can be used to extract faults signals algebraically as in \eqref{eq:algebraic_fault_estimate}. Note that, according to \eqref{eq:algebraic_fault_estimate}, the part of the extended state, $x_a$, that we use to reconstruct faults is $\bar{C}_a x_{a}$ with
	\begin{equation} 	\label{eq:c_bar_a}
		\bar{C}_a := {\left[\begin{array}{cccc}
				V_a^\top &
				\bar{C}_1^\top &
				\bar{C}_2^\top  &
				\bar{C}_3^\top
			\end{array}\right]^\top}.
	\end{equation}
	That is, we mainly require the observer to provide accurate estimates $\bar{C}_a \hat{x}_{a}$ of $\bar{C}_a x_{a}$ and not of the complete state $x_{a}$. Let us now define the estimation error
	\begin{equation*}
		e:=\hat{x}_a-x_a=z-x_a-E y=z-M x_a-E\nu.
	\end{equation*}
	Then, the estimation error dynamics is given by
	\begin{equation}
		\begin{aligned}
			\dot{e}= &N e+(N M+L C_a-M A_a) x_a+(G-M B_a) u \\
			&-M D_a \omega_{a} + (NE+L) \nu - E \dot{\nu}.
		\end{aligned}
		\label{eq:observer_dynamic}
	\end{equation}
	Given the algebraic relations in \eqref{eq:observer_matrices}, it can be verified that $G-MB_a = \mathbf{0}$, $N M+L C_a-M A_a = \mathbf{0}$, and $NE+L = K$. Therefore, \eqref{eq:observer_dynamic} amounts to
	\begin{equation}
		\begin{aligned}
			\dot{e} &= Ne - M D_a \omega_{a} + {\left[\begin{array}{cc}
					K & -E
				\end{array}\right]} {\left[\begin{array}{cc}
					\nu \\ \dot{\nu}
				\end{array}\right]}.
		\end{aligned}
		\label{eq:observer_dynamics_1}
	\end{equation}
	Define $\nu_a := {\left[\begin{array}{cc}
			\nu \\ \dot{\nu}
		\end{array}\right]}$, $e_d := \bar{C}_a e$ with $\bar{C}_a$ as in \eqref{eq:c_bar_a}, and
	\begin{equation}
		\bar{B} := {\left[\begin{array}{cc}
				K & -E
			\end{array}\right]}.
		\label{eq:b_bar}
	\end{equation}
	Then, the estimation error dynamics is given by
	\begin{equation}
		\left\{\begin{aligned}
			\dot{e} &=Ne - M D_a \omega_{a} + \bar{B} \nu_a, \\
			e_d &= \bar{C}_a e.
		\end{aligned}\right.
		\label{eq:error_system}
	\end{equation}
	Define the transfer matrices
	\begin{equation}
		\begin{aligned}
			T_{e_d \omega_a}(s) &:= -\bar{C}_a(s I-N)^{-1} M D_a, \\
			T_{e_d\nu_a}(s) &:= \bar{C}_a(s I-N)^{-1} \bar{B},
		\end{aligned}
		\label{eq:tfs}
	\end{equation}
	where $T_{e_d \omega_a}(s)$ and $T_{e_d\nu_a}(s)$ denote the corresponding transfer matrices from $\omega_a$ and $\nu_a$, both to $e_d$, respectively. Now, we can state the fault reconstruction problem that we seek to address.
	
	\begin{prob}\emph{\textbf{(Fault Reconstruction)} Consider the nonlinear system \eqref{eq:sys_transformed} with known input and output signals, $u(t)$ and $y(t)$, and let Assumptions \ref{assum:state_boundedness}-\ref{assum:noise_diff} be satisfied. Further, consider the approximated internal model \eqref{eq:ultra_local_internal_model}, the augmented dynamics \eqref{eq:augmented}, and the observer \eqref{eq:observer}. Design the observer gain matrices $(E,K)$ such that: \\ \textbf{1)} All trajectories of the error dynamics \eqref{eq:error_system} exist and are globally uniformly ultimately bounded in $t \geq 0$; \\ \textbf{2)} $\sup _{\mu \in \mathbb{R}^+} \sigma_{\max }(T_{e_d \omega_a}(i \mu)) < \lambda$, for some finite $\lambda > 0$,  transfer matrix $T_{e_d \omega_a}(i \mu)$ in \eqref{eq:tfs}, and frequencies $\mu \in \mathbb{R}^+$;\\ \textbf{3)} $\sqrt{\frac{1}{2 \pi} \operatorname{trace} \int_{-\infty}^{\infty} T_{e_d\nu_a}(i \mu) T_{e_d\nu_a}^{H}(i \mu) \mathrm{~d} \mu} < \gamma$, for some finite $\gamma > 0$, transfer matrix $T_{e_d\nu_a}(i \mu)$ in \eqref{eq:tfs}.}
	\end{prob} \label{prob}
	
	Under Assumptions \ref{assum:state_boundedness}-\ref{assum:noise_diff}, The essence of Problem 1 is to finding a fault estimator that: ensure a bounded estimation error, $e(t)$; the $H_{\infty}$-norm of $T_{e_d \omega_a}(s)$, the transfer matrix from external disturbances and model mismatches to the performance output $\bar{C}_a e$ (fault and some state estimation errors) is upper bounded by $\lambda$; the $H_2$-norm of $T_{e_d \nu_a}(s)$, the transfer matrix from measurement noise to the performance output, is upper bounded by $\gamma$; and, for $\omega_{a}(x, u, t) = \mathbf{0}$ and $\nu_{a}(t) = \mathbf{0}$, $e(t)$ goes to zero asymptotically (internal stability).

	\begin{rem} \emph{\textbf{($H_2$/$H_{\infty}$ Interpretation)}
			In the SISO case, the $H_{\infty}$-norm of a transfer function $T(s)$ can be interpreted as the highest peak value in its frequency response function. In other words, it is the largest amplification over all frequencies of input signals passed through $T(s)$ \cite[pp. 73]{scherer2000linear}. Then, minimizing the $H_{\infty}$-norm would decrease the worse-case amplification of inputs (disturbances) in the frequency domain. The $H_2$-norm of $T(s)$ has a stochastic interpretation -- it equals the asymptotic output variance of the system when driven by white noise signals \cite[pp. 74]{scherer2000linear}. Then, minimizing the $H_2$-norm would decrease the effect (in terms of variance) of stochastic inputs on the system output.}\linebreak 
	\end{rem}
	
	\section{Fault estimator design} \label{sec:lmi_based_design}
	
	\subsection{ISS Estimation Error Dynamics}
	This section outlines the process of deriving Linear Matrix Inequality (LMI) conditions for designing the matrices $E$ and $K$ of the augmented system observer \eqref{eq:observer}. As a stepping stone, we provide a sufficient condition for internal stability, which ensures asymptotic stability of the origin of the estimation error dynamics when the perturbations, $\omega_{a}$ and $\nu_{a}$, are both zero. Additionally, we prove the boundedness of the estimation error in the presence of perturbations using the concept of input-to-state stability, see \cite{sontag2008input}.
	
	\begin{defn} \emph{\textbf{(Input-to-State Stability)} The error dynamics \eqref{eq:observer_dynamics_1} is said to be Input-to-State Stable (ISS) if there exist a class $\mathcal{K} \mathcal{L}$ function $\psi(\cdot)$ and a class $\mathcal{K}$ function $\bar\zeta(\cdot)$ such that for any initial estimation error $e(t_0)$ and any bounded inputs $\omega_a(x, u, t)$ and $\nu_a(t)$, the solution $e(t)$ of \eqref{eq:observer_dynamics_1} exists for all $t \geq t_{0}$ and satisfies
			\begin{equation}\label{ISS_def}
				\|e(t)\| \leq \psi\left(\left\|e\left(t_{0}\right)\right\|, t-t_{0}\right) + \bar\zeta(\|
				{\left[\begin{array}{cc}
						\omega_a(x, u, t)  \\ \nu_a(t)
					\end{array}\right]}
				\|_{\infty}).
		\end{equation}}
		\label{def:iss}
	\end{defn}

	We remark that ISS of the error dynamics \eqref{eq:observer_dynamics_1} implies boundedness of the estimation error for bounded $\omega_{a}$ and $\nu_{a}$. This follows directly from $\eqref{ISS_def}$.

	The next proposition formalizes an LMI-based condition that guarantees an ISS estimation error dynamics \eqref{eq:observer_dynamics_1} with respect to input $[\omega_a^\top,\nu_a^\top]^\top$. 
	
	\begin{prop}\emph{\textbf{(ISS Estimation Error Dynamics)} Consider the error dynamics \eqref{eq:observer_dynamics_1} and suppose there exist matrices $P \in {\mathbb{R}^{n_z \times n_z}}, P \succ0, R \in {\mathbb{R}^{n_z \times m}}$, and $ Q \in {\mathbb{R}^{n_z \times m}}$ satisfying the inequality
			\begin{equation}
				X 	 +  \epsilon I \preceq 0,
				\label{eq:stability_lmi}
			\end{equation}
			for some given $\epsilon>0$ and matrix $X$ defined as
			\begin{equation}
				\begin{aligned}
					X :=&A_{a}^{\top}  P +A_{a}^{\top} C_{a}^{\top} R^{\top}-C_{a}^{T}  Q^{T} + P A_{a} +R C_{a} A_{a} -  Q C_{a}, 
				\end{aligned}
				\label{eq:X}
			\end{equation}
			with $A_a$ and $C_a$ in \eqref{eq:augmented_matrices}. Then, the estimation error dynamics in \eqref{eq:observer_dynamics_1} is ISS with input $[\omega_a^\top,\nu_a^\top]^\top$. Moreover, the ISS-gain from $[\omega_a^\top,\nu_a^\top]^\top$ to the estimation error $e$ in \eqref{eq:observer_dynamics_1} is upper bounded by $ 2 \|P {\left[\begin{array}{ccc} (I+E C_a) D_a & -K & E \end{array}\right]} \| \epsilon^{-1}$ with $E = P^{^{-1}} R $ and $K = P^{^{-1}} Q $.}
		\label{propos:stability}
	\end{prop}
	\emph{\textbf{Proof}}: The proof can be found in Appendix \ref{ap:propos1_proof}.
	\hfill $\blacksquare$
	

	
	
	Proposition 1 is employed to ensure that all trajectories of the estimation error dynamics \eqref{eq:observer_dynamics_1} exist and are bounded for all $t \geq 0$. If the observer gains $(E, K)$ satisfy the ISS LMI in \eqref{eq:stability_lmi}, boundedness and asymptotic stability for vanishing $\omega_a$ and $\nu_a$ is guaranteed. This follows directly from Assumptions \ref{assum:state_boundedness}-\ref{assum:noise_diff}, and Definition \ref{def:iss}.
	
	\subsection{$H_{\infty}$ Performance Criteria}
	To enhance the performance of fault reconstruction, our aim is to minimize the impact of $\omega_a$ (treated as an external disturbance) on the estimation error dynamics \eqref{eq:observer_dynamics_1}. We could use the ISS formulation in Proposition \ref{propos:stability} to formulate an optimization problem where we minimize the ISS gain while regarding the LMI in \eqref{eq:stability_lmi} as an optimization constraint.This approach can effectively reduce the effect of $\omega_a$ and $\nu_a$ on the complete vector of estimation errors $e$. However, it is essential to keep in mind that the purpose of the observer is to reconstruct faults only. Therefore, the performance of irrelevant estimation errors is not of direct concern. Accordingly, we intend to minimize the $H_{\infty}$-norm of the transfer function from $\omega_a$ to the relevant (according to \eqref{eq:algebraic_fault_estimate}) estimation errors $e_d$ as defined in \eqref{eq:tfs}.
	
	\begin{defn} \emph{\textbf{($H_{\infty}$-norm)}
			The $H_{\infty}$-norm $\|T_{e_d \omega_a}\|_{\infty}$ \linebreak of the transfer matrix $T_{e_d \omega_a}(s)$ is given by
			\begin{equation*}
				\|T_{e_d \omega_a}\|_{\infty} := \sup _{\mu \in \mathbb{R}^+} \sigma_{\max }(T_{e_d \omega_a}(i \mu)).
		\end{equation*}}
		\label{def:Hinf}
	\end{defn}

	The following proposition formalizes an LMI-based condition guaranteeing $\|T_{e_d \omega_a}\|_{\infty}<\lambda$, i.e., a finite $H_{\infty}$-norm of the transfer function $T_{e_d \omega_a}(s)$.
	
	\begin{prop}\emph{\textbf{(Finite $H_{\infty}$-norm)}
			Consider the \linebreak estimation error dynamics \eqref{eq:error_system} with transfer matrix $T_{e_d \omega_a}(s)$ in \eqref{eq:tfs}. Assume there exist matrices $P\succ0, R,$ $Q$, and scalar $\lambda > 0$ satisfying
			\begin{equation} \label{eq:hinf_lmi}
				\begin{aligned}
					&\left[\begin{array}{ccc}
						X & -(P + R C_a) D_a & \bar{C}_a^{\top} \\
						* & -\lambda I  & \boldsymbol{0} \\
						* & * & -\lambda I
					\end{array}\right] \prec 0,\\
				\end{aligned}
			\end{equation}
			with $X$ defined in \eqref{eq:X}, $\bar{C}_a$ in \eqref{eq:c_bar_a}, and the remaining matrices in \eqref{eq:augmented_matrices}. Then, $\|T_{e_d \omega_a}\|_{\infty}<\lambda$.}
		\label{propos:finite_Hinf}
	\end{prop}
		\emph{\textbf{Proof}}: The proof can be found in Appendix \ref{ap:propos2_proof}.
	\hfill $\blacksquare$
	
	\subsection{$H_{2}$ Performance Criteria}
	The other terms affecting the performance of the fault reconstruction are measurement noise and its derivative, collected in the vector $\nu_a$. We seek to minimize the effect of $\nu_a$ on the estimation error dynamics \eqref{eq:observer_dynamics_1}. To this end, we seek to minimize the $H_2$-norm of the transfer function from $\nu_a$ to the performance output $e_d$ defined in \eqref{eq:tfs}.
	
	\begin{defn} \emph{\textbf{($H_2$-norm)}
			The $H_2$-norm, $\|T_{e_d \nu_a}\|_{H_2}$, \linebreak of the transfer matrix $T_{e_d \nu_a}(s)$ is given by
			\begin{equation*}
				\|T_{e_d \nu_a}\|_{H_2} := \sqrt{\frac{1}{2 \pi} \operatorname{trace} \int_{-\infty}^{\infty} T_{e_d\nu_a}(i \mu) T_{e_d\nu_a}^{H}(i \mu) \mathrm{~d} \mu}.
		\end{equation*}}
		\label{def:H2}
	\end{defn}

	The following proposition formalizes an LMI-based condition guaranteeing $\|T_{e_d \nu_a}\|_{H_2}<\gamma$, i.e., a finite $H_2$-norm of $T_{e_d \nu_a}(s)$.
	\begin{prop}\emph{\textbf{(Finite $H_2$-norm)}
			Consider the \linebreak estimation error dynamics \eqref{eq:error_system} with transfer matrix $T_{e_d \nu_a}(s)$ in \eqref{eq:tfs}. Assume there exist matrices $P\succ0, R,$ $Q$, and scalar $\gamma > 0$ satisfying
			\begin{equation} \label{eq:h2_lmi}
				\begin{aligned}
					& \left[\begin{array}{cc}
						X &  {\left[\begin{array}{cc}
								Q & -R
							\end{array}\right]}\\
						* & -\gamma I
					\end{array}\right] \prec 0, \quad\left[\begin{array}{cc}
						P & \bar{C}_a^{\top} \\
						* & Z
					\end{array}\right] \succ 0,\quad
					\operatorname{trace}(Z)<\gamma,
				\end{aligned}
			\end{equation}
			with $X$ defined in \eqref{eq:X} and $\bar{C}_a$ in \eqref{eq:c_bar_a}. Then, \linebreak $\|T_{e_d \nu_a}\|_{H_2}<\gamma$.}
		\label{propos:finite_H2}
	\end{prop}
		\emph{\textbf{Proof}}: The proof can be found in Appendix \ref{ap:propos3_proof}. 
	\hfill $\blacksquare$
	
	Using Proposition \ref{propos:finite_Hinf}, we can formulate a semi-definite program where we seek to minimize the $H_{\infty}$-norm of $T_{e_d \omega_a}(s)$. Similarly, using Proposition \ref{propos:finite_H2}, we can formulate another semi-definite program where we seek to minimize the $H_2$-norm of $T_{e_d \nu_a}(s)$. However, in the presence of both unknown perturbations ($\omega_a$ and $\nu_{a}$), the $H_{\infty}$-norm and $H_2$-norm cannot be minimized simultaneously due to conflicting objectives. To attenuate the measurement noise effect on the estimations errors, a relatively slow observer is required, which does not react to every small change. In contrast, to reduce the effect of $\omega_a$, a high-gain observer is preferred, which tries to estimate the fault as accurately as possible. It follows that there is a trade-off between estimation performance the noise sensitivity. 
	
	To address this trade-off, a convex program where we seek to minimize the $H_{\infty}$-norm and constrain the $H_2$-norm,
	 for the same performance output $e_d$, is proposed. The same tools allow minimizing the $H_2$-norm for a constrained $H_{\infty}$-norm, for which we present simulation results. Moreover, we add the ISS LMI in \eqref{eq:stability_lmi} as a constraint to these programs to enforce that the resulting observer also guarantees boundedness for bounded perturbations and asymptotic stability for vanishing $\omega_a$ and $\nu_{a}$. The latter is essential to avoid observer divergence.
	
	\begin{thm}\emph{\textbf{(Fault Estimator Design)}
			Consider the augmented dynamics \eqref{eq:augmented}, the estimator \eqref{eq:observer}, the \linebreak corresponding estimation error dynamics \eqref{eq:error_system}, and the transfer functions \eqref{eq:tfs}. To design optimal mixed $H_2/H_{\infty}$ fault estimators \eqref{eq:observer}-\eqref{eq:algebraic_fault_estimate}, solve the following convex program
			\begin{equation}
				\begin{array}{cl}
					\min \limits_{P, R, Q, Z, \lambda, \gamma}  & \lambda \\
					\text{s.t.} & 						\vspace{1 mm}
					X 	 +  \epsilon I \preceq 0, \\
					&\left[\begin{array}{ccc}
						X & -(P + R C_a) D_a & \bar{C}_a^{\top} \\
						* & -\lambda I  & \boldsymbol{0} \\
						* & * & -\lambda I
					\end{array}\right] \prec 0,\\[5mm]
					& \left[\begin{array}{cc}
						X &  {\left[\begin{array}{cc}
								Q & -R
							\end{array}\right]}\\
						* & -\gamma I
					\end{array}\right] \prec 0, \quad\left[\begin{array}{cc}
						P & \bar{C}_a^{\top} \\
						* & Z
					\end{array}\right] \succ 0, \quad P \succ 0,\\[3mm]
					& \operatorname{trace}(Z)<\gamma, \quad \gamma \leq \gamma_{max}, \quad  \lambda, \gamma  > 0,
				\end{array}
				\label{eq:mimization}
			\end{equation}
			with given $\epsilon, \gamma_{max} > 0$, $\bar{C}_a$ in \eqref{eq:c_bar_a}, $X$ as defined in \eqref{eq:X}, and the remaining matrices as defined in \eqref{eq:augmented_matrices}. Denote the optimizers as $P^\star, R^\star, Q^\star, Z^\star, \lambda^\star, \gamma^\star$ and define the matrices $E^\star := P^{\star^{-1}} R^\star$ and $K^\star := P^{\star^{-1}} Q^\star$. Then, we have the following:
			\begin{enumerate}
				\item The ISS-gain from inputs $\omega_a$ and $\nu_a$ to the estimation error $e$ in \eqref{eq:error_system} is upper bounded by \linebreak $ 2 \|P^\star {\left[\begin{array}{ccc} (I+E^\star C_a) D_a & -K^\star & E^\star \end{array}\right]} \| \epsilon^{-1}$.
				\item The $H_{\infty}$-norm of transfer function $T_{e_d \omega_a}(s)$ in \eqref{eq:tfs} of the error dynamics \eqref{eq:error_system} is upper bounded by $\lambda^\star$ (i.e., $\|T_{e_d \omega_a}\|_{\infty}<\lambda^\star$).
				\item The $H_2$-norm of transfer function $T_{e_d \nu_a}(s)$ in \eqref{eq:tfs} of the error dynamics \eqref{eq:error_system} is upper bounded by $\gamma^\star$ (i.e., $	\|T_{e_d \nu_a}\|_{H_2}<\gamma^\star$).
		\end{enumerate}}
		\label{theorem:optimal_estimator}
	\end{thm}
	\emph{\textbf{Proof}:}
	Theorem \ref{theorem:optimal_estimator} follows from the above discussion and Propositions \ref{propos:stability}-\ref{propos:finite_H2}.
	\hfill $\blacksquare$
	\section{Simulation Results}  \label{sec:sim_results}
\begin{figure}[t]
		\centering
		\includegraphics[width = 0.45\textwidth]{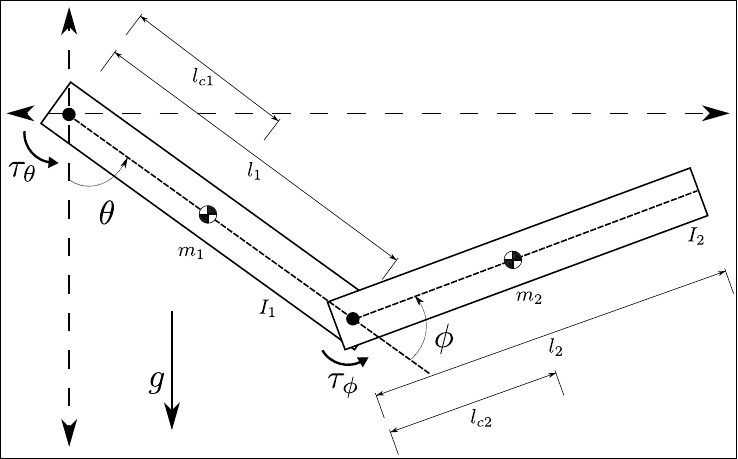}
		\caption{Schematic of the two-link robot manipulator.}
		\label{fig:PEM}
	\end{figure}

	In this section, the proposed method is evaluated by a two-link robotic manipulator with revolute joints, as shown in Fig. \ref{fig:PEM}. This system is employed as a case study due to its highly nonlinear behavior. The Euler-Lagrange equations of motions of this system are described by two second-order coupled differential equations (one for each link), which are given as \cite[pp. 86-88]{sira2018active}
	\begin{equation*}
		\underline{M}(q)\ddot{q} + 	\underline{C}(q,\dot{q})\dot{q} + \underline{G}(q) = \tau + \tau_f - \underline{D}\dot{q},
	\end{equation*}
	with
	\begin{equation*}
		\begin{split}
			&\underline{M}(q) = \begin{bmatrix} 	\underline{M}_{11} &  	\underline{M}_{12} \\
				* & 	\underline{M}_{22} \end{bmatrix}, \quad \underline{D} = \begin{bmatrix} d_1 & 0 \\ 0 & d_2 \end{bmatrix}, \quad 	\underline{G}(q)  = \begin{bmatrix} \underline{G}_1 \\ \underline{G}_2 \end{bmatrix},\\
			&\underline{C}(q,\dot{q}) =  \begin{bmatrix} -2m_2l_1l_{c_2}\sin(\phi) \dot{\phi} & -m_2l_1l_{c_2}\sin(\phi)\dot{\phi} \\
				m_2l_1l_{c_2}\sin(\phi)\dot{\theta} & 0\end{bmatrix}, \\
		\end{split}
	\end{equation*}
	where
	\begin{equation*}
		\begin{aligned}
			\underline{M}_{11} = &m_1l_{c_1}^2+m_2l_1^2  + m_2l_{c_2}^2 + 2m_2l_1l_{c_2}\cos(\phi) + I_1+I_2,\\
			\underline{M}_{12} = &m_2l_{c_2}^2 + m_2l_1l_{c_2}\cos(\phi) + I_2, \quad 	\underline{M}_{22} = m_2l_{c_2}^2+I_2,\\ 
			\underline{G}_1 = &\left( m_1l_{c_1}+m_2l_1 \right)g\sin(\theta)+m_2l_{c_2}g\sin(\theta + \phi), \\
			\underline{G}_2 = &m_2l_{c_2}g\sin(\theta + \phi),
		\end{aligned}
	\end{equation*}
	and $q = [\theta \,\, \phi]^T$ is the generalized coordinate, $\tau =[\tau_{\theta} \,\, \tau_{\phi}]^T = [0.5 \sin\left(\frac{2\pi}{40}t\right) \,\, 0]^T$ the input, and $\tau_f = [\tau_{f_{\theta}} \,\, \tau_{f_{\phi}}]^T$ the actuator fault. In addition, the measurement is $q$, which contains the angular positions of both links. The parameters are as follows \cite[p. 92]{sira2018active}:
	\begin{equation*}
		\begin{array}{lll}
			m_1 = 0.263 \hspace{1 mm} kg, & m_2 = 0.1306 \hspace{1 mm} kg,   \\
			I_1 = 0.002 \hspace{1 mm} Nm^2, & I_2 = 0.00098 \hspace{1 mm} Nm^2,  \\
			l_1 = 0.3 \hspace{1 mm} m, & l_2 = 0.3 \hspace{1 mm} m, \\
			l_{c_1} = 0.15 \hspace{1 mm} m, & l_{c_2} = 0.15 \hspace{1 mm} m , \\
			d_1 = 0.03 \hspace{1 mm} Nms/rad, & d_2 = 0.005 \hspace{1 mm} Nms/rad, \\
			g = 9.81 \hspace{1 mm} m/s^2.
		\end{array}
	\end{equation*}

			\begin{figure}[t!] 	
		\centering
		\smallskip
		\includegraphics[width=.95\linewidth,keepaspectratio]{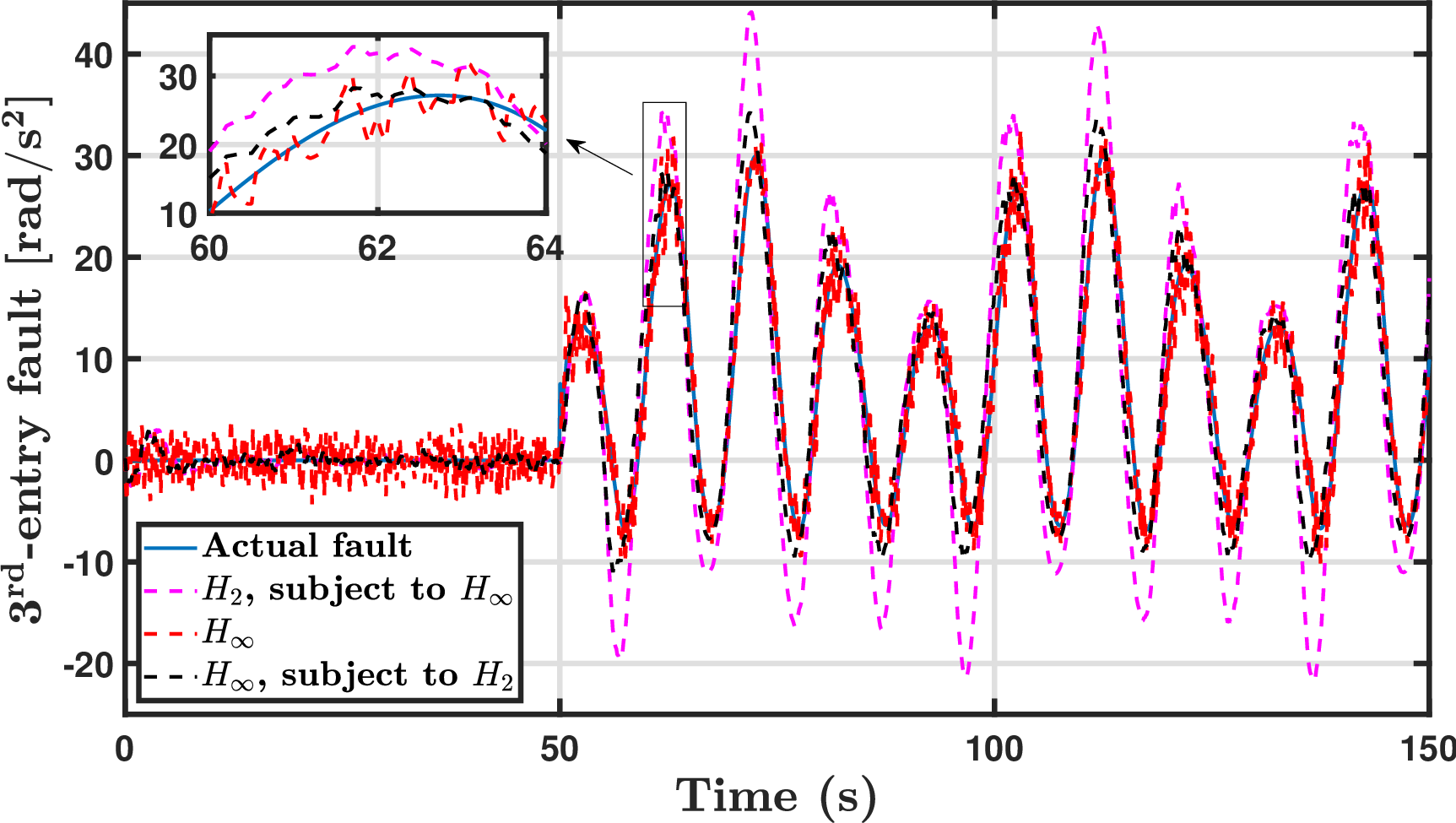}
		\caption{The actual third entry fault and its estimates.}
		\label{fig:estimated_fault_1}
	\end{figure}

	By selecting $x_1 :=q$ and $x_2 :=\dot{q}$, the system can be written directly in the form of \eqref{eq:sys_transformed} (no transformation is needed for this example) as follows
	\begin{equation} \label{eq:simulation_ss}
		\left\{\begin{aligned}
			\dot{x} =& A x +S \big(g(V x, u)+ f_n(x)\big), \\
			y=& C x + \nu,
		\end{aligned}\right.
	\end{equation}
	where $x := [x_1, x_2]^T = [q, \dot{q}]^T$ is the state vector,
	\begin{equation*}
		\begin{aligned}
			A &= \left[\begin{array}{cc}
				\boldsymbol{0}_{2 \times 2} & I_2 \\
				\boldsymbol{0}_{2 \times 2} & -\underline{M}_l^{-1}\underline{D} \\
			\end{array}\right],
			&S&=\left[\begin{array}{l}
				\boldsymbol{0}_{2 \times 2} \\
				I_2
			\end{array}\right], \\
			C&=\left[\begin{array}{cc}
				I_2 & \boldsymbol{0}_{2 \times 2}
			\end{array}\right],
			&V&= I_4, \\
		\end{aligned}
	\end{equation*}
	\begin{equation*}
		\begin{aligned}
			g(V x, u) = &\underline{M}^{-1}(x_1)\big(u - \underline{D} x_2-\underline{C}(x_1,x_2)x_2 - \underline{G}(x_1)\big) \\
			&+ \underline{M}_l^{-1}\underline{D}x_2, \quad u = \tau, \quad f_n(x) = \underline{M}^{-1}(x_1) \tau_f,
		\end{aligned}
	\end{equation*}
	and $\underline{M}_l = \underline{M}(\boldsymbol{0})$. The measurement noise $\nu$ is generated from a uniform distribution with the maximum of $0.1$ $rad$ and the minimum of $-0.1$ $rad$, sample time of $0.1$ $sec$, for both sensors, which is a large noise level. We have used such noise levels here to compare different above-mentioned approaches. We set the initial conditions of the system to zero. The actuator fault signal is given as
	\begin{equation*}{\label{eq:fault_scenaro}}
		\tau_f(t) = \begin{cases}
			\tau_{f_{\theta}} = 0, & \quad \text{if } t<50, \\
			\tau_{f_{\phi}} = 0, & \quad \text{if } t<50, \\
			\tau_{f_{\theta}} = 0.2\sin\left(\frac{2\pi}{10}(t-50)\right), & \quad \text{if } t\geq50, \\
			\tau_{f_{\phi}} = -0.05, & \quad \text{if } t\geq50. \\
		\end{cases}
	\end{equation*}
	Note that the nonlinear fault $f_n(x, u)$ has two fault entries; one affects the third entry and the other, the fourth entry of the manipulator dynamics. Although each actuator fault affects just one actuator, since we are estimating $f_n(x)$ and not $\tau_f$, the occurrence of only one actuator fault causes non-zero values for both entries of $f_n(x)$.

%

			\begin{figure}[t!] 	
		\centering
		\smallskip
		\includegraphics[width=.95\linewidth,keepaspectratio]{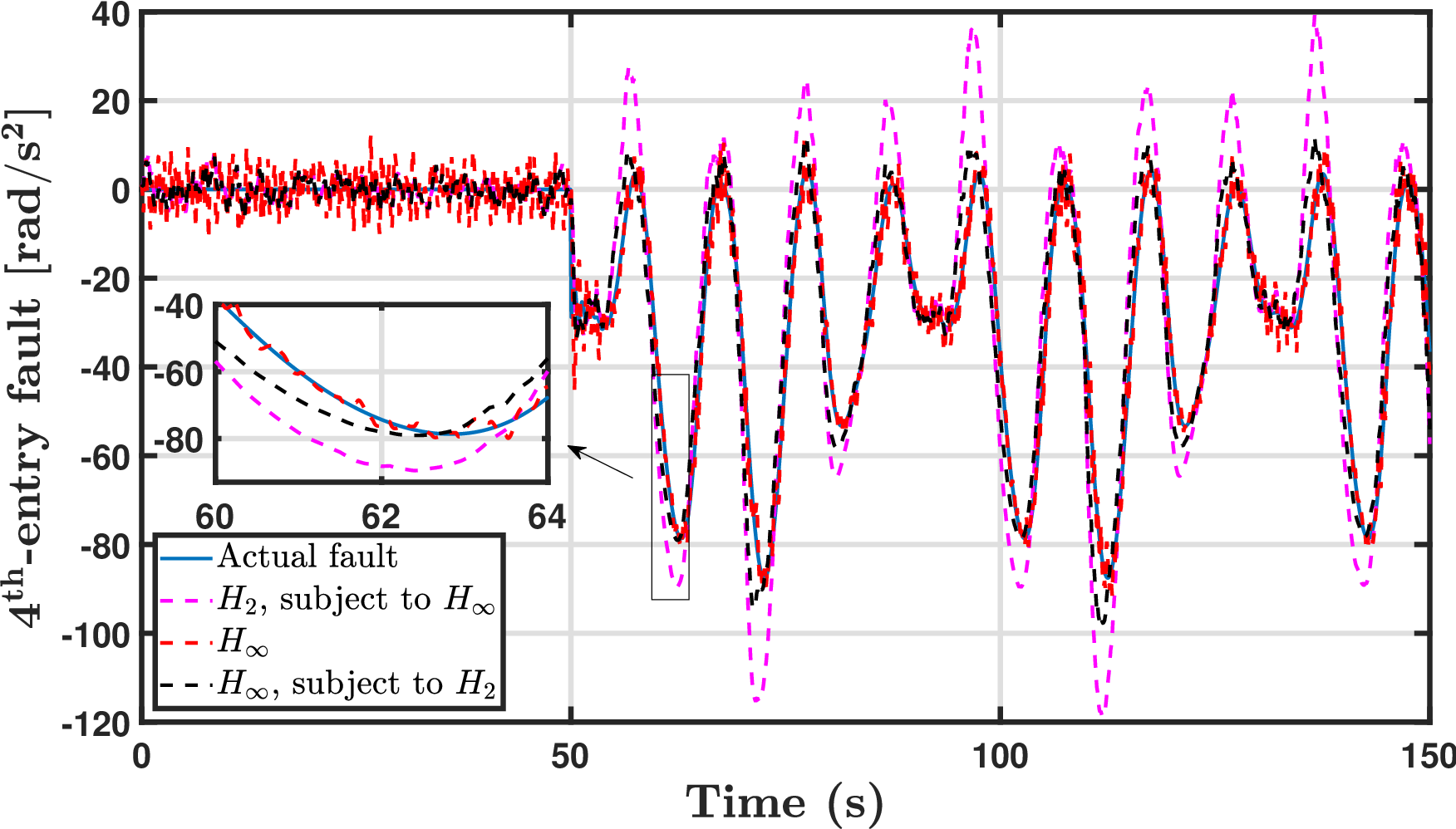}
		\caption{The actual forth entry fault and its estimates.}
		\label{fig:estimated_fault_2}
	\end{figure}
	
	Next, we augment the system \eqref{eq:simulation_ss} using \eqref{eq:augmented} with $r=4$. For the augmented system, the observer of the form \eqref{eq:observer} can be designed in three different ways as follows:
	\begin{enumerate}
		\item Only minimizing the $H_{\infty}$-norm of $T_{e_d \omega_a}(s)$ in \eqref{eq:tfs} of the error system \eqref{eq:error_system}, subject to ISS LMI in \eqref{eq:stability_lmi}.
		\item Minimizing the $H_{\infty}$-norm of $T_{e_d \omega_a}(s)$ in \eqref{eq:tfs} of the error system \eqref{eq:error_system}, subject to ISS LMI in \eqref{eq:stability_lmi} and an upper bound for the $H_2$-norm of transfer function $T_{e_d \nu_a}(s)$ in \eqref{eq:tfs} of the error system \eqref{eq:error_system} (Theorem \ref{theorem:optimal_estimator}).
		\item Minimizing the $H_2$-norm of transfer function $T_{e_d \nu_a}(s)$ in \eqref{eq:tfs} of the error system \eqref{eq:error_system}, subject to ISS LMI in \eqref{eq:stability_lmi} and an upper bound for the $H_{\infty}$-norm of $T_{e_d \omega_a}(s)$ in \eqref{eq:tfs} of the error system \eqref{eq:error_system} (i.e., opposite of the previous one).
	\end{enumerate}
	The designed observer is estimating the lumped disturbance and the nonlinear actuator fault estimates can be computed using \eqref{eq:algebraic_fault_estimate}. The initial conditions of the observers in the simulation are taken as a vector with all values equal to $0.01$. Figure \ref{fig:estimated_fault_1} and Figure \ref{fig:estimated_fault_2} depict the estimated nonlinear faults and their actual values, using all the observers described above for the faults affecting the third and the fourth entries of the manipulator dynamics, respectively. It can be seen that only for the $H_{\infty}$-norm is minimized, although the fault estimation performance is decent, the results are too noisy. Therefore, we minimize the $H_{\infty}$ and constrain the $H_{2}$ norm or the other way around (known as a mixed $H_2/H_{\infty}$ approach). Herein, the fault estimation performance and the noise attenuation are always a trade-off, and the proposed synthesis approach allows us to make this trade-off in a constructive manner.

	

	\section{Conclusion} \label{sec:conclusion}
	This paper presents a method for the robust reconstruction of time-varying faults in nonlinear systems, using ultra-local observers for modeling faults and nonlinearities. The design approach proposes semi-definite programs that allow for a trade-off between fault estimation performance and noise sensitivity. The numerical simulations for a two-link manipulator illustrate the performance of the proposed approach and the potential for highly nonlinear systems. Future work includes incorporating model uncertainty.

	\appendix

\section{Proof of Proposition 1}  \label{ap:propos1_proof}  
Let us first introduce the following lemma, which is used to ensure ISS using an ISS Lyapunov function.
\begin{lem}\emph{\textbf{(ISS Lyapunov Function~{\cite[Thm. 4.19]{khalil2002nonlinear}})}
		Consider the error dynamics \eqref{eq:observer_dynamics_1} and let $W(e)$ be a continuously differentiable function such that
		\begin{equation*}
			\alpha_{1}(\|e\|) \leq W(e) \leq \alpha_{2}(\|e\|),
		\end{equation*}
		\begin{equation*}
			\dot{W}(e) \leq-W_{3}(e), \quad \hspace{-1mm} \forall \hspace{1mm}\|e\| \geq \xi\biggl(\biggl\| {\left[\begin{array}{cc}
					\omega_a(x, u, t)  \\ \nu_a(t)
				\end{array}\right]} \biggl\|\biggl),
		\end{equation*}
		where $\alpha_{1}(\cdot)$ and $\alpha_{2}(\cdot)$ are class $\mathcal{K}_{\infty}$ functions, $\xi(\cdot)$ is a class $\mathcal{K}$ function, and $W_{3}$ is a continuous positive definite function. Then, the estimation error dynamics \eqref{eq:observer_dynamics_1} is ISS with gain $ \bar{\zeta} = \alpha_{1}^{-1}(\alpha_{2}(\xi))$.}
	\label{lem: iss}
\end{lem}

Let $W(e):={e}^{T} P {e}$ be an ISS Lyapunov function candidate. Then, it follows from \eqref{eq:observer_dynamics_1} that
\begin{equation}
	\begin{aligned} \dot{W}(e) &\leq e^{T} \Delta e  - 2 e^{T} P
		\bar{B}_a
		{\left[\begin{array}{cc}
				\omega_a(x, u, t)  \\ \nu_a(t)
			\end{array}\right]},
	\end{aligned}
	\label{eq:lyapanov}
\end{equation}
where
\begin{equation}
	\begin{aligned} \Delta := &N^{T} P+P N, \qquad
		\bar{B}_a := &{\left[\begin{array}{cc}
				M D_a & -\bar{B}
			\end{array}\right]}.
	\end{aligned}
	\label{eq:delta}
\end{equation}
Inequality \eqref{eq:lyapanov} implies the following 
\begin{equation}
	\begin{aligned} \dot{W}(e) \leq& -\lambda_{\min }(-\Delta) \|e\|^{2} + 2 \|e\| \|P \bar{B}_a\| \biggl\|{\left[\begin{array}{cc}
				\omega_a  \\ \nu_a
			\end{array}\right]}\biggl\| \\
		=& - (1-\theta) \lambda_{\min }(-\Delta) \|e\|^{2} - \theta \lambda_{\min }(-\Delta) \|e\|^{2} \\
		&+ 2 \|e\| \|P \bar{B}_a\|\biggl\|{\left[\begin{array}{cc}
				\omega_a  \\ \nu_a
			\end{array}\right]}\biggl\|,
	\end{aligned}
	\label{eq:iss_analysis}
\end{equation}
for any $\theta \in (0,1)$. Therefore, by \eqref{eq:iss_analysis} and Lemma \ref{lem: iss}, if $\Delta$ is negative definite, error dynamics \eqref{eq:observer_dynamics_1} is ISS with inputs $\omega_{a}$ and $\nu_{a}$ and linear ISS-gain
\begin{equation}
	\bar\zeta\biggl(\biggl\|{ \left[\begin{array}{cc}
			\omega_a  \\ \nu_a
		\end{array}\right]}\biggl \|\biggl)  = \frac{2 \|P \bar{B}_a\|}{\theta \lambda_{\min }(-\Delta)}\biggl\|{\left[\begin{array}{cc}
			\omega_a  \\ \nu_a
		\end{array}\right]}\biggl\|.
	\label{eq: gamma}
\end{equation}
Without loss of generality, for numerical tractability, we enforce $\Delta + \epsilon I \preceq 0$ for some arbitrarily small given $\epsilon >0$ instead of $\Delta \prec 0$.

Now, we need to show that $\Delta$ is equivalent to \eqref{eq:X}. Using $\Delta$ defined in \eqref{eq:delta} and \eqref{eq:observer_matrices}, we can write $\Delta$ in terms of the original observer gains $(E,K)$ as
\begin{equation*}
	\begin{aligned}
		\Delta = 	&A_{a}^{T}\left(I+E C_{a}\right)^{T} P-C_{a}^{T}  K P+P\left(I+E C_{a}\right) A_{a} \\
		&-P K C_{a},
	\end{aligned}
\end{equation*}
Consider the following change of variables
\begin{equation}
	R :=P E,  \qquad	Q :=P K.
	\label{eq:variable_change}
\end{equation}
Applying \eqref{eq:variable_change} on the above expanded $\Delta$, the linear inequality \eqref{eq:X} can be concluded. Clearly, $\Delta + \epsilon I \preceq 0$ implies $\lambda_{\min }(-\Delta) \geq \epsilon$. Then, using \eqref{eq: gamma}, we can conclude the bound on the ISS-gain.

\section{Proof of Proposition 2}  \label{ap:propos2_proof}  
Let us first introduce the following lemma, in which we state a necessary and sufficient condition for having a bounded $H_{\infty}$-norm of $T_{e_d \omega_a}(s)$.
\begin{lem} \emph{\textbf{(Finite $H_{\infty}$-norm~\cite[Prop. 3.12]{scherer2000linear})}
		Consider the estimation \linebreak error dynamics \eqref{eq:error_system} with transfer matrix $T_{e_d \omega_a}(s)$ in \eqref{eq:tfs}. \linebreak Assume $N$ is Hurwitz and consider a finite $\lambda > 0$. Then, the following statements are equivalent
		\begin{enumerate}
			\item $\|T_{e_d \omega_a}\|_{\infty}<\lambda$.
			\item There exists a  $P \in {\mathbb{R}^{n_z \times n_z}}, P \succ 0$ satisfying
			\begin{equation} \label{eq:hinf_not_lmi}
				\left[\begin{array}{ccc}
					N^{\top} P+P N & -P M D_a & \bar{C}_a^{\top} \\
					* & -\lambda I & \boldsymbol{0} \\
					* & * & -\lambda I
				\end{array}\right] \prec 0,
			\end{equation}
			with $(N,M)$ in \eqref{eq:observer_matrices}, $\bar{C}_a$ in \eqref{eq:c_bar_a}, and $D_a$ in \eqref{eq:augmented_matrices}.
	\end{enumerate}}
	\label{lem:finite_Hinf}
\end{lem}
By applying the change of variables in \eqref{eq:variable_change}, one can see that \eqref{eq:hinf_lmi} is equivalent to \eqref{eq:hinf_not_lmi}, which based on \ref{lem:finite_Hinf} concludes the result of the proposition.

\section{Proof of Proposition 3}  \label{ap:propos3_proof}  
Let us first introduce the following lemma, in which we state a necessary and sufficient condition for having a bounded $H_2$-norm of $T_{e_d \nu_a}(s)$.
\begin{lem} \emph{\textbf{(Finite $H_2$-norm~\cite[Prop. 3.13]{scherer2000linear})}
		Consider the \linebreak estimation error dynamics \eqref{eq:error_system} with transfer matrix $T_{e_d \nu_a}(s)$ in \eqref{eq:tfs}. Assume $N$ is Hurwitz and consider a finite $\gamma > 0$. Then, the following statements are equivalent
		\begin{enumerate}
			\item $	\|T_{e_d \nu_a}\|_{H_2}<\gamma$.
			\item There exists a  $P \succ 0$ and $Z$ satisfying
			\begin{equation} \label{eq:h2_not_lmi}
				\begin{aligned}
					& \left[\begin{array}{cc}
						N^{\top} P+P N & P \bar{B}\\
						* & -\gamma I
					\end{array}\right] \prec 0, \quad\left[\begin{array}{cc}
						P & \bar{C}_a^{\top} \\
						* & Z
					\end{array}\right] \succ 0,
					\\
					& \operatorname{trace}(Z)<\gamma,
				\end{aligned}
			\end{equation}
			with $N$ in \eqref{eq:observer_matrices}, $\bar{B}$ in \eqref{eq:b_bar}, and $\bar{C}_a$ in \eqref{eq:c_bar_a}.
	\end{enumerate}}
	\label{lem:finite_H2}
\end{lem}
By applying the change of variables in \eqref{eq:variable_change}, one can see that \eqref{eq:h2_lmi} is equivalent to \eqref{eq:h2_not_lmi}, which based on \ref{lem:finite_H2} concludes the result of the proposition.


	\bibliography{refs}             
	
	
	
	
	
	
	

\end{document}